\documentclass[letterpaper]{article} 
\usepackage{aaai2026}  
\usepackage{times}  
\usepackage{helvet}  
\usepackage{courier}  
\usepackage[hyphens]{url}  
\usepackage{graphicx} 
\usepackage{natbib}  
\usepackage{caption} 
\usepackage{algorithm}
\usepackage{algorithmic}
\usepackage{multirow} 
\usepackage{amsmath} 
\usepackage{booktabs}

\usepackage{newfloat}
\usepackage{listings}
\DeclareCaptionStyle{ruled}{labelfont=normalfont,labelsep=colon,strut=off} 
\floatstyle{ruled}
\newfloat{listing}{tb}{lst}{}
\floatname{listing}{Listing}
\title{SRLF: An Agent-Driven Set-Wise Reflective Learning Framework for Sequential Recommendation}
\author {
    Jiahao Wang\textsuperscript{\rm 1},
    Bokang Fu\textsuperscript{\rm 1},
    Yu Zhu\textsuperscript{\rm 1},
    Yuli Liu\textsuperscript{\rm 1}\thanks{*Corresponding author}
}
\affiliations {
    \textsuperscript{\rm 1}Qinghai University\\
    wangjiahao2025@gmail.com, fubokang1123@gmail.com, zhuyu@qhu.edu.cn, liuyuli012@gmail.com
}

\usepackage{bibentry}

\begin{document}

\maketitle

\begin{abstract}
LLM-based agents are emerging as a promising paradigm for simulating user behavior to enhance recommender systems. However, their effectiveness is often limited by existing studies that focus on modeling user ratings for individual items. This point-wise approach leads to prevalent issues such as inaccurate user preference comprehension and rigid item-semantic representations.

To address these limitations, we propose the novel Set-wise Reflective Learning Framework (SRLF). Our framework operationalizes a closed-loop “assess-validate-reflect” cycle that harnesses the powerful in-context learning capabilities of LLMs. SRLF departs from conventional point-wise assessment by formulating a holistic judgment on an entire set of items. It accomplishes this by comprehensively analyzing both the intricate interrelationships among items within the set and their collective alignment with the user's preference profile. This method of set-level contextual understanding allows our model to capture complex relational patterns essential to user behavior, making it significantly more adept for sequential recommendation. Extensive experiments validate our approach, confirming that this set-wise perspective is crucial for achieving state-of-the-art performance in sequential recommendation tasks.
\end{abstract}

\section{Introduction}
In today's information-rich online environments, recommender systems have emerged as a cornerstone technology, essential for mitigating information surplus and facilitating the discovery of personalized content \cite{li2024recommender}. The field's trajectory is marked by an evolution of modeling paradigms, from early Markov Chain \cite{cheng2013you, he2016fusing} and RNN \cite{beutel2018latent, rakkappan2019context, xu2019recurrent} to the subsequent rise of Transformers \cite{kang2018self}. The recent advent of LLMs \cite{achiam2023gpt} has opened a new frontier. LLM-based agents \cite{shen2023hugginggpt, yao2023react}, endowed with sophisticated capabilities in reasoning, planning, and nuanced decision-making, are now emerging as a powerful paradigm for simulating human behavior. This paradigm shift leverages natural language to offer a crucial advantage: the ability to articulate the logic behind recommendations, thereby enhancing transparency and user trust. This development holds significant promise for engineering the next generation of intelligent and adaptive recommender systems. Despite the advent of these powerful new tools, the fundamental recommendation paradigm has not evolved accordingly, failing to fully leverage the contextual reasoning capabilities now at our disposal.

Our research is motivated by the fundamental question of whether LLMs can uncover more complex and profound dependencies when processing sequential information at a set level, as opposed to the traditional item-level. We hypothesize that by providing an entire set of items as a unified input, we can leverage the powerful contextual modeling and semantic integration capabilities of LLMs to capture latent structural information and interaction patterns that are missed by item-by-item analysis. This set-level modeling approach is poised to unlock the holistic comprehension capabilities of LLMs, yielding richer and more effective information representation for complex recommendation tasks.

However, the prevailing paradigm has historically centered on modeling user preferences from isolated, point-wise interactions with individual items \cite{nithish2024efficient, kehan2025precise}. Such a perspective, while effective for simple scenarios, fundamentally struggles to comprehend the rich, contextual dynamics of user behavior in complex environments. These systems fail to grasp the contextual nature of recommendation. On one hand, they oversimplify user preferences into a static collection of individual likes \cite{sun2024beyond}; on the other, they treat items as isolated entities, overlooking the complex inter-item dynamics that truly shape user choice. To illustrate this limitation, consider a user browsing music albums: their preference for a jazz album may depend not only on their historical interest in jazz, but also on what other albums are simultaneously available in the candidate set: the presence of classical music might enhance the appeal of jazz through contrast, while an abundance of similar jazz albums might lead to choice paralysis or preference for diversity. This contextual interdependence among candidate items represents a fundamental aspect of human decision-making that point-wise models systematically ignore. Consequently, there is a clear need for frameworks that explicitly model both intra-set dependencies and their holistic alignment with user preferences.

Nowhere are these limitations more consequential than in sequential recommendation \cite{li2017neural}. The very nature of this task, predicting the next item in a user's interaction trajectory, demands a holistic comprehension of the order, context, and interplay among preceding items \cite{tang2018personalized}. Moreover, in sequential contexts, the interdependencies extend beyond simple pairwise relationships to encompass higher-order patterns across multiple items. A user's next choice is influenced not just by their last interaction, but by the collective narrative and thematic coherence of their recent activity sequence. This temporal set-level coherence represents a critical signal that traditional point-wise methods cannot capture. This requisite for contextual understanding \cite{yang2021context} exposes a fundamental flaw in traditional models: their isolated, point-wise perspective limits their ability to capture the complex, higher-order relational patterns that constitute a sequence's meaning. While Transformer-based models represent a significant leap forward in capturing pair-wise item relations \cite{wan2022cross}, they function primarily as opaque mechanisms. This operational opacity stems from a core conceptual limitation: the absence of a dedicated framework designed to explicitly reason about the holistic, emergent properties of an entire item set. This critical gap underscores the urgent need for a new paradigm: one that moves beyond both atomistic point-wise assessments and opaque pair-wise modeling to embrace a truly holistic, context-aware perspective.

In response to these challenges, we put forward the Set-wise Reflective Learning Framework (SRLF), a novel paradigm that advances the principles of set-level learning \cite{liu2022determinantal, liu2024learning, liu2024universal} by harnessing the sophisticated capabilities of LLM agents for a new class of personalized recommendations. Specifically, our framework reconceptualizes the assessment process. Transcending the limitations of both point-wise and pair-wise assessments, SRLF formulates a holistic judgment over the candidate set as a cohesive whole.

This is achieved through a closed-loop “assess-validate-reflect” reasoning process that empowers the agent to continuously refine its understanding of user profiles and item semantics, where the crucial assess step is accomplished by systematically weighing two key aspects: the internal, inter-item dynamics within the set, and the set's external alignment with the user's preference profile. This set-level contextual understanding, empowered by the potent in-context learning of LLMs, enables our model to discern higher-order relational patterns (\textit{e.g.}, collaborative and sequential dependencies) that are crucial for modeling realistic user choices but are challenging for traditional models to capture. By reasoning at the set level, our approach naturally aligns with the sequential nature of the task, making it more effective at navigating the nuanced demands of sequential recommendation.

The main contributions of this work are as follows:
\begin{itemize}
\item We propose the SRLF, a novel paradigm for LLM agents that operationalizes a closed-loop “assess-validate-reflect” reasoning cycle to transcend the limitations of traditional point-wise recommendation.
\item We formulate a novel set-level reasoning methodology. By holistically analyzing inter-item relationships, this methodology enables the capture of complex contextual patterns crucial for modeling user behavior in sequential recommendation.
\item We conduct extensive experiments that validate the superiority and adaptability of our set-wise approach, establishing a new state-of-the-art in performance and confirming the criticality of this perspective.
\end{itemize}

\section{Related Work}
\paragraph{Sequential Recommendation Models.} The fundamental task of sequential recommendation (SR) is to predict the relevance of subsequent items based on a user's historical interaction sequence, a challenge predominantly addressed by leveraging sequential models. Early approaches, such as Markov Chain (MC) \cite{cheng2013you, he2016fusing}, model this process under the assumption that the next action depends only on the most recent items. To overcome the limited memory of MC and capture more complex sequence patterns, subsequent research turned to deep learning. Among these, Convolutional Neural Network (CNN) conceptualizes the embedding sequence as an image to which convolutional operators are applied \cite{tang2018personalized, yan2019cosrec}, while Recurrent Neural Network (RNN) leverages their inherent sequential architecture to capture longer-range dependencies \cite{beutel2018latent, rakkappan2019context, xu2019recurrent}. Self-attention based models have recently emerged as the dominant paradigm in SR, a trend initiated by the seminal work of \cite{kang2018self}. Building upon this pioneering work, subsequent studies have focused on augmenting the attention mechanism with sequence order information \cite{li2020time, li2023edge} and contextual signals \cite{huang2018csan, wu2020deja} to further improve performance. Hybrid models that fuse self-attention with other powerful sequential architectures have been extensively explored to enhance SR performance. Furthermore, these approaches capitalize on the respective advantages of each component, integrating techniques such as contrastive learning \cite{qiu2022contrastive, xie2022contrastive}, GNN \cite{xu2019graph, yang2022multi}, CNN \cite{chen2022double, jiang2023adamct}, and RNN \cite{xia2017attention, li2021lightweight}.

The research paradigm in SR has remained largely confined to item-level transitions, neglecting the investigation of preferences from a more holistic, set-level perspective.

\paragraph{Agent-based Recommender Systems.} The emergence of LLM-based agents has introduced a transformative paradigm to recommender systems. This has led to the development of agent-based recommender systems, which are characterized by their ability to autonomously process user-item interactions and leverage sophisticated reasoning for personalization, thereby opening up several distinct research directions. One line of research focuses on ranking-oriented agents\cite{wang2023recmind, zhao2024let, wang2024macrec}, which infer user preferences from historical behavior to generate recommendations. This approach is exemplified by RecMind\cite{wang2023recmind}, a unified LLM agent for direct recommendation generation. In parallel, a second direction explores simulation-oriented agents\cite{zhang2024generative, zhang2024agentcf} that leverage role-playing capabilities to simulate human-like behavior. A key example is AgentCF\cite{zhang2024agentcf}, which simulates user-item interactions through agent-based collaborative filtering. A third research direction explores interactive conversational agents\cite{shu2024rah,  huang2025recommender} that frame recommendation as dialogue-based intent understanding. This approach is exemplified by RAH \cite{shu2024rah}, a framework designed for iterative user preference refinement within a human-recommender interaction loop.

Despite their diverse methodologies, these existing works predominantly operate at a point-wise level, thus overlooking the important challenge of set-level recommendation.

\paragraph{Temporal Sets Prediction.} In contrast to widely studied time series forecasting \cite{wu2021autoformer, zhou2021informer} or sequence prediction \cite{bengio2015scheduled, chiu2018state}, the field of Temporal Sets Prediction (TSP) is specifically concerned with the complex interplay of intra-set item correlations and inter-set temporal dependencies to predict future sets. The seminal work of \cite{hu2019sets2sets} first formalized the task of TSP and proposed a novel set-to-set framework. To capture dependencies across temporal sets, their approach leverages well-established sequential models, including multi-head self-attention \cite{sun2020dual} and RNN. Recently, some studies have sought to model the intricate relationships between sets by propagating information, exemplified by \cite{yu2020predicting}, which applies graph convolutions on a dynamic graph of these relationships to learn comprehensive set representations. The training of TSP is also constrained by their reliance on item-level loss functions (e.g., weighted mean square loss \cite{hu2019sets2sets} and binary cross-entropy loss \cite{sun2020dual, yu2022element}), which are incapable of capturing the complicated relationships among sets.

While TSP frameworks represent an important step toward modeling temporal set dynamics, their reliance on traditional sequential architectures and item-level losses limits their ability to incorporate sophisticated reasoning about set-level interdependencies. These limitations inspire our agent-driven approach, which leverages LLMs to perform holistic set-wise assessments and reflective learning. Building on TSP's emphasis on inter-set and intra-set relationships, we introduce an adaptive framework that explicitly addresses mismatch through a closed-loop mechanism, enabling more nuanced capture of user preferences in sequential contexts.

\section{Methodology}

\begin{figure*}[h]
	\centering
	\includegraphics[width=0.8\linewidth]{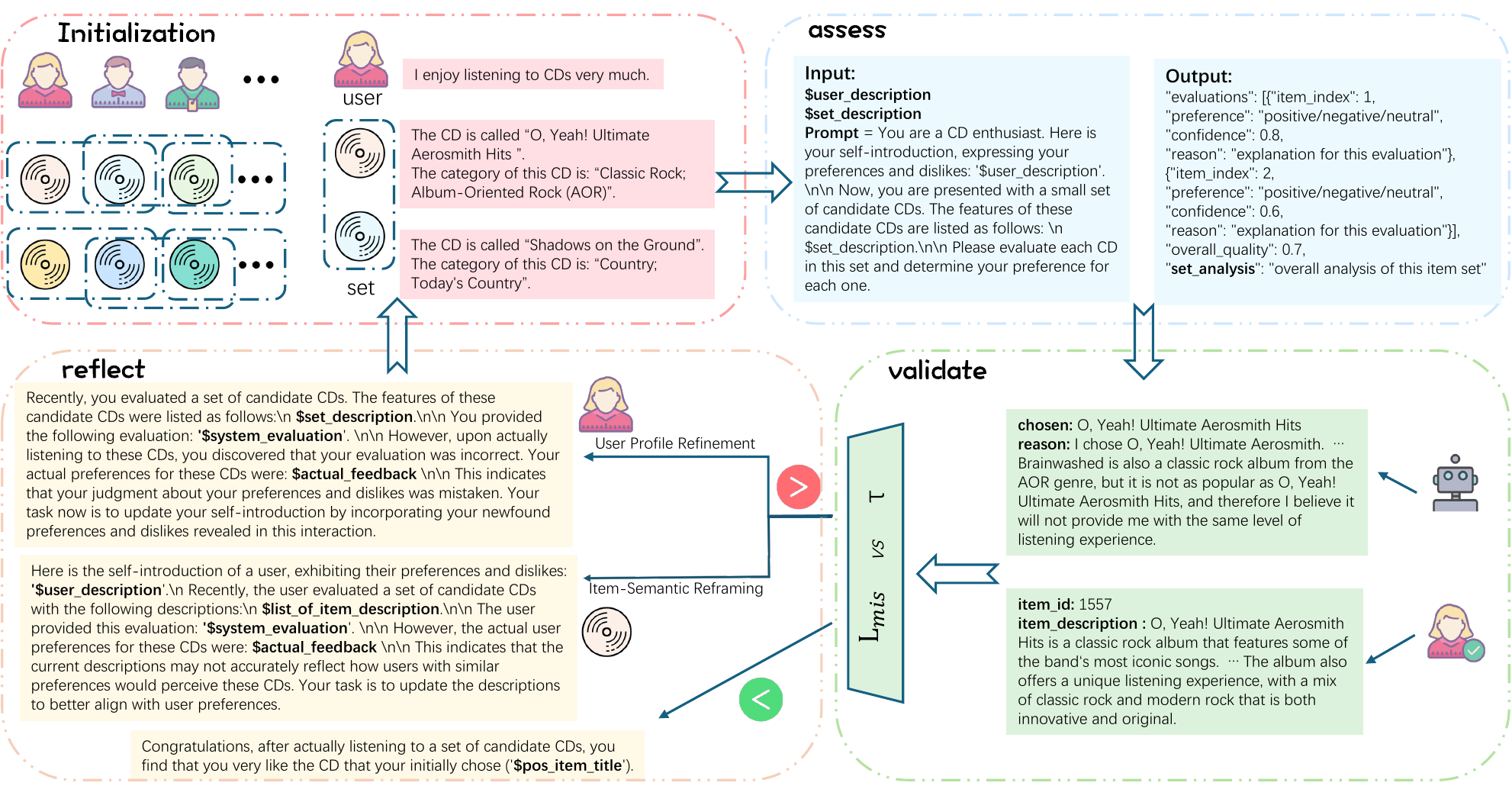}
        \caption{The Set-wise Reflective Learning Framework (SRLF). The framework introduces a closed-loop learning mechanism. It begins with a set-wise assessment of items by an LLM agent. Mismatches with user feedback then trigger a dual-path reflection, which concurrently refines both the user profile  and item semantics for continuous adaptation.}
	\label{fig:overall}
\end{figure*}

In this section, we present the proposed \textbf{S}et-wise \textbf{R}eflective \textbf{L}earning \textbf{F}ramework, named \textbf{SRLF}. The overall architecture of our framework is illustrated in Figure 1. It comprises three core components: the Set-wise Assessment Agent (SAA), the Validation via Set-wise Mismatch Loss, and the Dual-Path Reflective Learning mechanism. We begin by establishing the preliminaries of recommendation with Large Language Models and highlight the limitations of conventional methodologies, before delving into a detailed exposition of each component.

\subsection{Preliminaries}

In a conventional recommender system, the primary objective is to model user preferences based on a set of historical interactions $\mathcal{D} = \{(u, i)\}$, where $u \in \mathcal{U}$ represents a user and $i \in \mathcal{I}$ an item \cite{bottou2010large}. The goal is typically to learn a preference function $f(u, i)$ that predicts the user's affinity for an item.

Recent advancements have seen Large Language Models (LLMs) being leveraged for this task. Prevailing LLM-based paradigms operationalize recommendation on a \textit{point-wise} basis, where an agent assesses items in isolation by computing $f_{point-wise}(P_u, i_k)$ for each $i_k \in I_s$ independently \cite{kehan2025precise}. Within our agent-driven framework, both the user's preference profile ($P_u$) and the attributes of each item ($D_i$) are instantiated as natural language texts. The user profile $P_u$ serves as a textual summary of the user's tastes and historical behaviors, while the item's attributes $D_i$ form a semantic description of its characteristics. This textual representation is crucial as it allows the LLM-based agent to directly leverage its sophisticated language understanding and reasoning capabilities for preference modeling. Despite its effectiveness, this approach fundamentally struggles to comprehend the rich, contextual dynamics of user choice, as it overlooks the intricate inter-item relationships within a candidate set. For instance, a user's preference for one item is often contingent on the other available options.

To transcend these limitations, our work reconceptualizes the recommendation process, shifting from isolated, point-wise evaluations to a holistic, \textit{set-level} judgment. In contrast to the point-wise formulation, our set-wise paradigm considers the entire candidate set as a unified input by computing $f_{set-wise}(P_u, I_s)$. This fundamental shift from $(P_u, i_k)$ to $(P_u, I_s)$ enables the agent to produce a comprehensive assessment that considers both individual item attributes and their collective interplay. To effectively capture fine-grained inter-item relationships, we further propose an overlapping set partitioning strategy that divides the candidate items into smaller, overlapping subsets for granular evaluation. This set-wise paradigm, which we detail in the subsequent sections, enables the modeling of more complex and realistic user decision-making processes.

\subsection{Set-wise Assessment Agent}

The primary objective of the Set-wise Assessment Agent (SAA) is to assess whether a given set of items, denoted as \(I_s = \{i_1, i_2, \dots, i_n\}\), aligns with a user's latent preferences.In a departure from conventional point-wise assessment methodologies, the SAA formulates a holistic judgment by performing a comprehensive analysis of the intricate interrelationships among items within the set, in addition to their collective alignment with the user's preference profile, \(P_u\). By scrutinizing these inter-item correlations, the model can capture complex relational patterns that are fundamental to user behavior in sequential contexts, thereby making the approach more suitable for the task of sequential recommendation. This process is actuated by an LLM, which drives the agent's exploration into the relational landscape of the item set. Leveraging a precisely formulated prompt, the agent is tasked not merely with passive evaluation but with an active investigation of inter-item dynamics to elicit its output in a predefined structured format.

To enable more nuanced assessment of inter-item relationships, we introduce an overlapping set partitioning mechanism. Given a candidate set $I_s = \{i_1, i_2, \dots, i_n\}$, we partition it into overlapping subsets of size $k$ (typically $k=2$):

\begin{equation} 
    \mathcal{S} = \left\{ \{S_j | S_j = \{i_j, i_{j+1}, \dots, i_{j+k-1}\} \mid 1 \leq j \leq n-k+1 \right\}.
\end{equation}
For example, given items $\{1,2,3,4,5,6\}$, we create overlapping pairs: $\{(1,2), (2,3), (3,4), (4,5), (5,6)\}$. This overlapping strategy ensures that each item appears in multiple subsets, allowing the model to evaluate its relationships with different neighboring items.

In realistic recommendation scenarios, this overlapping strategy offers distinct advantages over alternative grouping methods, such as non-overlapping partitions or holistic full-set evaluation. Compared to non-overlapping partitions (e.g., $\{(1,2), (3,4), (5,6)\}$), our approach preserves local contextual integrity. Non-overlapping methods would arbitrarily sever crucial sequential links, for instance, the relationship between items 2 and 3 would be entirely ignored, despite its potential importance in a user's decision-making process. Conversely, when compared to evaluating the entire set at once, the overlapping subset approach provides a more granular and structured signal for learning. While a holistic assessment captures the overall set quality, it often fails to pinpoint which specific inter-item dynamics contributed to the user's final decision. Our method deconstructs this complex evaluation into manageable and localized assessments, and produces fine-grained feedback, which is essential for the subsequent dual-path reflection mechanism to effectively refine user profiles and item semantics. Furthermore, this multi-context evaluation, where each item is assessed with different neighbors, enhances the robustness of the preference signals by mitigating the impact of noise from any single assessment.

For each subset $S_j$, the SAA performs individual assessment to capture fine-grained inter-item relationships:

\begin{equation} 
    A_{llm}^{(j)} = f_{eval}(P_u, S_j; \theta_{prompt}),
\end{equation}
where $A_{llm}^{(j)}$ represents the assessment result for subset $S_j$, $P_u$ represents the user's preference profile, and $\theta_{prompt}$ denotes the parameters governing the LLM's assessment template. Each assessment provides both item-level preference predictions and subset-level compatibility scores. The overlapping nature ensures that the compatibility of each project can be evaluated in multiple environments, so as to generate more reliable evaluation signals and inform subsequent training and reasoning processes. 

\subsection{Validation via Set-wise Mismatch Loss}
The assessments rendered by the SAA represent the agent's preliminary judgments based on its current understanding of the user and items. However, for the framework to learn and adapt, these predictive hypotheses must be systematically benchmarked against actual user responses. We therefore introduce a formal validation stage to quantify the discrepancy between the agent's predictions and the ground truth, thereby generating a concrete error signal to guide the subsequent reflective learning phase.

Following the initial assessment rendered by the SAA, the framework initiates a validation process. This crucial step involves juxtaposing the agent's assessment ($A_{llm}$) with the ground truth feedback ($F_{truth}$) to formally quantify the divergence between the predicted and actual user response. To enable a more fine-grained quantification of this discrepancy, we formulate a novel Set-wise Mismatch Loss, denoted as $\mathcal{L}_{mis}$. This loss is specifically engineered to encapsulate the magnitude of the deviation in the agent's set-level judgment.

Departing from conventional binary loss functions that provide a coarse-grained signal of correctness, our loss function is engineered to offer a more nuanced measure of predictive error. Specifically, it quantifies the aggregate preference discrepancy between the LLM's assessment and the user's true preferences across the entire set, thereby enabling a holistic evaluation of the agent's judgment. For overlapping set partitions, the loss is computed as:

\begin{equation}
    \mathcal{L}_{mis} = \sum_{j=1}^{n-k+1} \sum_{i \in S_j} |A_{llm}^{(j)}(i)_{pref} - A_{true}(i)_{pref}|.
\end{equation}
This loss function aggregates the absolute errors across all overlapping subsets within the candidate item set. For each overlapping subset $S_j$, it calculates the divergence between the LLM's assigned preference score for subset $j$, $A_{llm}^{(j)}(i) {pref}$, and the ground truth score, $A {true}(i)_{pref}$, derived from user feedback for each item $i$ within that subset. By summing these individual deviations across all overlapping subsets, the formulation provides a comprehensive measure of the agent's predictive misalignment that accounts for the overlapping evaluation structure. The overlapping nature ensures that each item's mismatch is captured multiple times across different subset contexts, providing a more robust and context-aware error signal.

When the computed \(\mathcal{L}_{mis}\) exceeds a predefined threshold \(\tau\), the system generates a mismatch signal, indicating a significant deviation in the LLM's assessment that invokes the subsequent reflective learning mechanism for correction. The threshold \(\tau\) can be dynamically adjusted based on the user's historical interaction patterns and the specific requirements of the recommendation scenario.

\subsection{Dual-Path Reflective Learning}

Upon detecting a discrepancy signal, the SRLF initiates a dual-path reflective mechanism. This reflective process is particularly crucial for set-level learning due to the inherent complexity of the assessment. Unlike point-wise errors, a set-level mismatch can stem from multiple sources: an inaccurate user profile, a flawed understanding of inter-item relationships (e.g., synergy or contrast), or a misinterpretation of the set's overall thematic coherence. The reflective mechanism is essential for diagnosing the specific cause of the error and guiding a targeted correction, thereby making the sophisticated task of set-level preference modeling tractable and adaptive.

This mechanism operationalizes the agent-driven exploration at the heart of our framework. Rather than passively receiving a gradient signal to update weights, the agent actively explores the error space by investigating the source of the predictive mismatch. This exploration is structured along two distinct dimensions to rectify the model's internal biases through a concurrent analysis of the discrepancy. This mechanism is predicated on the key insight that a predictive mismatch stems from two primary sources: a flawed model of the user's preferences, or a mischaracterization of the item's semantics. Inspired by human cognitive processes of post-hoc error analysis, our framework employs two concurrent, complementary paths to disambiguate and address the root cause of the error.

The first path, User Profile Refinement, targets inaccuracies in the user preference model, while the second, Item-Semantic Reframing, corrects suboptimal representations of item semantics. By operating in parallel, this dual-path approach ensures a holistic correction, distributing the adaptive responsibility instead of incorrectly attributing the entire error to a single component. This leads to more stable and effective learning.

The reflective learning process is enhanced by the overlapping set structure, which provides multiple perspectives on each item's assessment. We categorize the reflection scenarios based on the composition of positive and negative examples within each subset:

\begin{itemize}
\item \textbf{Two Positive Items}: When a subset contains two ground-truth positive items, the reflection focuses on understanding synergistic preferences.
\item \textbf{Two Negative Items}: When a subset contains two ground-truth negative items, the reflection emphasizes identifying common rejection patterns.
\item \textbf{Mixed Composition}: When a subset contains one positive and one negative item, the reflection analyzes contrastive preferences to better distinguish user likes from dislikes.
\end{itemize}

User Profile Refinement: The first path of our reflective mechanism addresses inaccuracies in the user preference model. In the event of discrepancies between the model's assessments across multiple overlapping subsets (${A_{llm}^{(j)}}$) and the corresponding ground truth feedback (${F_{truth}^{(j)}}$), the framework infers that the existing user profile ($P_u$) is suboptimal. Consequently, the LLM is prompted to perform a reflective analysis of these divergences across all evaluated subsets to generate a refined user profile, $P_u'$. We formalize this user profile refinement as a function that aggregates insights from multiple subset evaluations:

\begin{equation}
    P_u' = f_{r\_user}(P_u, \{A_{llm}^{(j)}\}, \{F_{truth}^{(j)}\}).
\end{equation}
This iterative refinement allows the model to more effectively track and adapt to the dynamic nature of user interests while leveraging the rich contextual information provided by the overlapping set structure.

Item-Semantic Reframing: The second reflective path addresses discrepancies that stem from suboptimal representation of item semantics. Specifically, when negative user feedback across multiple overlapping subsets (${F_{truth}^{(j)} | i \in S_j}$) contradicts the model's positive assessments (${A_{llm}^{(j)} | i \in S_j}$) for item $i$ appearing in various subsets $S_j$, the framework hypothesizes that the issue may lie not in user preference, but in how the item's attributes are presented in its description ($D_i$). To mitigate this, the LLM is prompted to reframe the description by analyzing the item's performance across all subsets where it appears, generating a new version $D_i'$ that aligns more closely with the user's cognitive perspective as evidenced by the collective feedback patterns. We formalize this process as:
\begin{equation}
    D_i' = f_{r\_item}(D_i, \{A_{llm}^{(j)} | i \in S_j\}, \{F_{truth}^{(j)} | i \in S_j\}).
\end{equation}
This multi-context reframing enables more robust item representation learning by considering how each item performs across different neighboring contexts.

\subsection{Overall Algorithmic Procedure of SRLF}
To synthesize the components delineated in the preceding subsections, we now formalize the complete procedural workflow of our Set-wise Reflective Learning Framework (SRLF). Given a user's historical interaction data and a candidate item set, the procedure commences with the Set-wise Assessment Agent (SAA) performing a holistic evaluation across overlapping subsets to generate an initial assessment. Subsequently, the Set-wise Mismatch Loss is computed to quantify the discrepancy between the agent's predictions and the ground truth feedback. This validation step generates an error signal that informs the core of our framework: the dual-path reflective learning phase. Within this phase, the framework leverages the mismatch signals to concurrently refine both the user profile and the item-semantic representations. This iterative, closed-loop mechanism ensures that the model continuously adapts and improves its understanding of user preferences and item characteristics.

\section{Experiments}
In this section, we present a series of experiments designed to thoroughly evaluate the performance of our proposed SRLF framework. We first introduce the experimental setup, then present the overall performance comparison against state-of-the-art baselines, and finally conduct an in-depth ablation study to isolate and analyze the contributions of SRLF's core components.
\begin{table*}[t]
\small
  \centering
    \begin{tabular}{lccccccccc} 
    \toprule
     \multirow{2}{*}{Method} 
     & \multicolumn{3}{c}{CDs$_\text{sparse}$} 
     & \multicolumn{3}{c}{CDs$_\text{dense}$}
     & \multicolumn{3}{c}{MovieLens} \\  
     \cmidrule(lr){2-4} \cmidrule(lr){5-7} \cmidrule(lr){8-10}  
     & NDCG@1 & NDCG@5 & NDCG@10 & NDCG@1 & NDCG@5 & NDCG@10 & NDCG@1 & NDCG@5 & NDCG@10 \\  
    \midrule
    BM25  & 0.0800 & 0.3066 & 0.4584 & 0.0600 & 0.2624 & 0.4325 & 0.0980 & 0.2102 & 0.4067 \\
    BPR   & 0.1300 & 0.3597 & 0.4907 & 0.1300 & 0.3485 & 0.4812 & 0.0960 & 0.2994 & 0.4543 \\
    SASRec & 0.1900 & 0.3948 & 0.5308 & 0.1300 & 0.3151 & 0.4676 & 0.1128 & 0.3346 & 0.4742 \\
    LLMRank & 0.1367 & 0.3109 & 0.4715 & 0.1333 & 0.3689 & 0.4946 & 0.1180 & 0.3773 & 0.4917 \\
    AgentCF & 0.1900 & 0.3466 & 0.5019 & 0.2067 & 0.4078 & 0.5328 & 0.1720 & 0.3903 & 0.4966 \\
    SRLF & \textbf{0.2400} & \textbf{0.4115} & \textbf{0.5478} & \textbf{0.2300} & \textbf{0.4552} & \textbf{0.5594} & \textbf{0.1780} & \textbf{0.4173} & \textbf{0.5266} \\
    \bottomrule
    \end{tabular}%
    \caption{Overall performance comparison on CDs sparse, CDs dense and MovieLens datasets.}
    \label{tab:performance_comparison}
\end{table*}
\subsection{Experimental Setup}
\textbf{Datasets.} We experimented with two benchmark datasets: the Amazon review dataset \cite{ni2019justifying}: “CDs and Vinyl”, and the MovieLens dataset \cite{harper2015movielens}. For our experimental setup, we constructed the evaluation datasets by sampling users from each benchmark. For the Amazon dataset, we randomly sampled 100 users, thereby minimizing the overhead of API calls. For the MovieLens dataset, we randomly sampled 500 users to ensure a representative subset. To investigate the impact of data sparsity on SR, for the Amazon dataset, we randomly sample both a dense (CDs dense) and a sparse (CDs sparse) subset to facilitate evaluation across diverse interaction scenarios. The MovieLens dataset is sampled following established practices. This allows us to test the robustness of our model under different data availability conditions.\\
\textbf{Evaluation.} We evaluate the performance using Normalized Discounted Cumulative Gain (NDCG@K), with K set to 1, 5, and 10. We adopt the leave-one-out evaluation protocol. For each user, the chronologically last item in their interaction sequence is held out as the ground-truth. The model is then tasked with ranking this target item against a set of nine items, randomly sampled from the item corpus, which follows prior popular works \cite{zhang2024agentcf, liu2025agentcf++}.\\
\textbf{Baseline.} To comprehensively situate the performance of SRLF, we compare the proposed method against a diverse suite of baseline methods spanning multiple paradigms:
\begin{itemize}
    \item \textbf{BM25} \cite{robertson2009probabilistic}: A classic retrieval algorithm that ranks candidates based on their textual similarity with user historical interactions.
    \item \textbf{BPR} \cite{rendle2012bpr}: A matrix factorization method that learns user and item representations by optimizing the BPR loss.
    \item \textbf{SASRec} \cite{kang2018self}: A Transformer-based model that captures sequential patterns from user historical interactions.
    \item \textbf{LLMRank} \cite{hou2024large}: A zero-shot ranker that leverages ChatGPT to score items, conditioned on user sequential interaction histories.
    \item \textbf{AgentCF} \cite{zhang2024agentcf}: An approach that simulates user-item interactions through collaborative learning between user-agents and item-agents.
\end{itemize}

\subsection{Overall Performance}

As shown in Table 1,  the overall performance comparison of our method against the baselines on the sampled datasets. The best-performing results for each metric are highlighted in \textbf{bold}. Our method is always better than the baseline of all evaluation scenarios, thereby highlighting the effectiveness of set-level learning for simulating personalized agents. 

Among the baseline models, we observe a distinct hierarchy. The traditional matrix factorization method BPR establishes a reasonable performance baseline but is clearly limited by its inability to capture sequential dependencies. The Transformer-based model SASRec significantly surpasses BPR, especially on the sparse dataset. This highlights the crucial role of modeling sequential patterns in user behavior. The agent-based model AgentCF delivers competitive results, performing on par with SASRec and even slightly better on the dense dataset. This indicates the potential of agent-based simulation. However, the fact that SRLF consistently outperforms all these approaches demonstrates that its set-level learning paradigm captures more complex user dynamics than existing paradigms.

The tuning-free methods demonstrate limited efficacy in our experiments. The BM25 model, while considering content, is constrained by its reliance on shallow lexical similarity and misses the deeper semantic and sequential patterns in user behavior. Most tellingly, even the advanced LLMRank, which prompts ChatGPT with user histories, significantly underperforms against trained models. This result, which aligns with findings in \cite{liu2023chatgpt}, underscores a critical challenge: the universal knowledge of large language models does not readily translate to an effective understanding of domain-specific user behavior patterns. These results collectively confirm that heuristic and zero-shot approaches are insufficient for this task, reinforcing the necessity for specialized, trainable architectures.

\subsection{Ablation Study}

\begin{table}[t]
\small
  \centering
    \begin{tabular}{lcccc}
    \toprule
     \multirow{2.5}{*}{Method}
     & \multicolumn{3}{c}{CDs$_\text{dense}$} \\     \cmidrule(lr){2-4} 
     & NDCG@1 & NDCG@5 & NDCG@10 \\    \midrule
    w/o Set-wise Assessment  & 0.1500 & 0.3811 & 0.4965 \\
    w/o Reflection Learning  & 0.1500 & 0.3498 & 0.5009 \\
    SRLF  & \textbf{0.2300} & \textbf{0.4552} & \textbf{0.5594} \\
    \bottomrule
    \end{tabular}%
    \caption{Ablation study.}
\end{table}

To validate the individual contributions of the key components in our proposed SRLF, we conducted an ablation study on two variants of our model. For this analysis, we use the dense dataset, as its richer interaction data provides a clearer setting to demonstrate the impact of our model's components. We designed two specific variants:
\begin{itemize}
    \item \textbf{SRLF w/o Set-wise Assessment}: In this variant, we remove the set-wise assessment module and replace it with a more conventional approach where the agent evaluates items individually, thus modeling preferences from a point-wise perspective.
    \item \textbf{SRLF w/o Reflection Learning}: This variant removes the dual-path reflective learning mechanism. The model performs only the set-wise assessment, without the subsequent refinement of user profiles and item semantics.
\end{itemize}

The results, presented in Table 2, reveal that the performance of SRLF drops significantly when either the set-wise assessment module or the reflective learning mechanism is removed. This demonstrates that both components are integral to the model's effectiveness.

Specifically, the \textit{SRLF w/o Set-wise Assessment} variant exhibits a notable performance degradation compared to the full SRLF model. This result validates our core hypothesis that a set-level perspective is superior to conventional point-wise approaches for sequential recommendation. By reasoning about the candidate set as a cohesive whole, SRLF can capture higher-order, contextual relationships among items that are missed by methods that assess items in isolation. This holistic understanding is essential for modeling the complex decision-making processes inherent in user sequential behavior.

Furthermore, the performance of \textit{SRLF w/o Reflection Learning} is also substantially lower than that of the full SRLF model across all metrics. This highlights the critical role of the reflective learning mechanism. By iteratively refining user profiles and item semantics based on feedback, the model can continuously adapt and correct its internal representations, leading to a more accurate understanding of user preferences. Without this mechanism, the model's learning is static and less capable of resolving discrepancies between its predictions and user behavior.

\section{Conclusion and Future Work}
In this work, we propose the Set-wise Reflective Learning Framework (SRLF), a novel paradigm designed to transcend the limitations of traditional point-wise evaluation in recommender systems. Our framework operationalizes a closed-loop ``assess-validate-reflect'' reasoning cycle, empowering LLM agents to formulate holistic judgments over entire sets of items. This approach affirms our central hypothesis that allowing an LLM to process items as a cohesive set, rather than individually, is critical for uncovering the deeper, structural dependencies inherent in user behavior. By comprehensively analyzing the intricate inter-item dynamics and their collective alignment with user preferences, SRLF effectively captures complex, high-order relational patterns that remain elusive to conventional models.

Extensive experiments conducted across multiple datasets with varying sparsity demonstrate the superiority and adaptability of our approach. The proposed SRLF consistently and significantly outperforms state-of-the-art baselines, including AgentCF, across all evaluated metrics. These results not only establish a new state-of-the-art performance benchmark for sequential recommendation but also underscore the criticality of adopting a set-wise perspective for a more nuanced comprehension of user behavior.

For future work, we identify two promising directions. First, we will explore adapting the SRLF framework to explicitly model and enhance recommendation diversity. Second, we will focus on optimizing the set-wise reflection mechanism to enhance computational efficiency, thereby facilitating its deployment in real-world industrial systems. We believe this agent-driven, set-centric paradigm offers a promising pathway toward engineering the next generation of intelligent and adaptive recommender systems.

\bibliography{aaai2026}


\end{document}